\documentclass[11pt]{scrartcl}
\usepackage{amsmath}
\usepackage{amssymb}
\usepackage{amsthm}
\usepackage{slashed}
\usepackage{graphicx}
\usepackage{color}
\usepackage[centering,headsep=15pt,head=80pt,foot=20pt,margin=60pt]{geometry}
\usepackage[scaled]{helvet}
\usepackage[slantedGreek]{mathptmx}

\newcommand{\Group}[2]{{ \hbox{{\itshape{#1}}($#2$)} }}
\newcommand{\U}[1]{\Group{U\kern0.05em}{#1}}
\newcommand{\SU}[1]{\Group{SU\kern0.1em}{#1}}
\newcommand{\SL}[1]{\Group{SL\kern0.05em}{#1}}
\newcommand{\Sp}[1]{\Group{Sp\kern0.05em}{#1}}
\newcommand{\SO}[1]{\Group{SO\kern0.1em}{#1}}

\newcommand{\mybar}[1]%
    {{\kern 0.8pt\overline{\kern -0.8pt#1\kern -0.8pt}\kern 0.8pt}}
\newcommand{\roughly}[1]%
    {{ \mathrel{\raise.3ex\hbox{ $#1$\kern-.75em\lower1ex\hbox{$\sim$}} } }}

\newcommand{\avg}[1]{\langle #1 \rangle}
\newcommand{\Avg}[1]{\left\langle #1 \right\rangle}
\newcommand{\nop}[1]{:\kern-.3em#1\kern-.3em:}

\providecommand{\abs}[1]{\lvert#1\rvert}

\newcommand{\del}{\partial}

\newcommand{\al}{\ensuremath{\alpha}}

\newcommand{\Ga}{\ensuremath{\Gamma}}
\newcommand{\de}{\ensuremath{\delta}}

\newcommand{\ep}{\ensuremath{\epsilon}}

\renewcommand{\th}{\ensuremath{\theta}}

\newcommand{\rh}{\ensuremath{\rho}}
\newcommand{\si}{\ensuremath{\sigma}}

\newcommand{\ch}{\ensuremath{\chi}}

\newcommand{\om}{\ensuremath{\omega}}

\newcommand{\n}{\notag \\}

\newcommand{\pfrac}[2]{\left(\frac{#1}{#2}\right)}

\numberwithin{equation}{section}
\numberwithin{figure}{section}

\begin{document}
\begin{titlepage}
\begin{flushright}
OU-HET-949 \\
\end{flushright}
\vskip 5em
\begin{center}
{\Large \bfseries
 Probing new intra-atomic force with isotope shifts\\
}
\vskip 4em
Kyoko Mikami$^\flat$,
Minoru Tanaka$^\flat$, \footnote{email: \texttt{tanaka@phys.sci.osaka-u.ac.jp}}
Yasuhiro Yamamoto$^\sharp$ \footnote{email: \texttt{yamayasu@yonsei.ac.kr}}
\vskip 4em
$^\flat$
  \textit{Department of Physics, Graduate School of Science, Osaka University, Toyonaka, Osaka 560-0043, Japan}\\
$^\sharp$
	\textit{Department of Physics and IPAP, Yonsei University, Seoul 03722 Republic of Korea}\\
\vskip 4em
\textbf{Abstract}
\end{center}
\medskip
\noindent

In the development of atomic clocks, some atomic transition frequencies are measured with remarkable precision.
These measured spectra may include the effects of a new force mediated by a weakly interacting boson. 
Such effects might be distilled out from possible violation of a linear relation in isotope shifts between two transitions, as known as King's linearity, with relatively suppressed theoretical uncertainties.
We discuss the experimental sensitivity to a new force in the test of the linearity as well as the linearity violation owing to higher-order effects within the Standard Model.
The sensitivity to new physics is limited by such effects.
We have found that, for Yb$^+$, the higher-order effect is in the reach of future experiments.
The sensitivity to a heavy mediator is also discussed.
It is analytically clarified that the sensitivity becomes weaker than that in the literature.
Our numerical results of the sensitivity are compared with other weak force search experiments.

\bigskip
\vfill
\end{titlepage}
\tableofcontents
\section{Introduction}
One of the main targets of the intensity frontier in particle physics is a new force carrier which is much lighter than the weak scale and very weakly interacts with the Standard Model particles.
Such a new particle is examined by low energy experiments and stellar observations. 

Among various experiments, measurements of transition frequencies in the development of atomic clocks achieve remarkable precision.
Their relative error is expected to be the order of $10^{-18}$ for Yb ion in the near future~\cite{Phys.Rev.Lett.116.063001}.
Therefore, precision atomic spectroscopy can be considered as a sensitive probe to a new force other than the Coulomb interaction.
In contrast to the extreme precision of the experiments, the theoretical calculation of atomic spectra suffers from uncertainties of the many-body system.

A possible way to reduce theoretical uncertainties is to utilize the isotope shift.
The atomic spectra differ from one isotope to another.
The shifts are considerably small, typically O(GHz), compared with the major part of the Coulomb interaction.
The remaining shifts are ascribed to two origins.
One is the mass shift and the other is the field shift.
The mass shift is caused by the modification of the kinetic term.
The field shift depends on the modification of the potential which originates from the nuclear charge distribution differences of isotopes.
At the leading order of these effects, the isotope shifts of two different transitions satisfy a linear relation.
This linear relation is called King's linearity~\cite{RefKing}.

The electron-neutron interaction given by a new force carrier also contributes the isotope shift.
In general, this effect violates King's linearity.
Using this property, the possibility to detect the contributions of the Higgs and $Z$ bosons is studied in Ref.~\cite{1601.05087}.
As revisited in the next section, their effects are highly suppressed by the large scale differences between the electroweak and the atomic physics.
However, the violation of King's linearity could be still a good probe of a new force mediated by a boson lighter than about 1 MeV~\cite{1704.05068,PhRva.D96.015011,Delaunay:2017dku}.

In this paper, we study the violation of King's linearity by a new force which is mediated by a light new boson. 
In particular, its sensitivity to the new physics is discussed by not only numerical but also analytic means.
We also study the next-leading-order contribution of the field shift, which also violates the linearity~\cite{RefFs,1709.00600}.
We argue how the field-shift non-linearity limits the experimental sensitivity to the new physics.
The expected bounds on the new force in future experiments are compared with the other known constraints.
It turns out that King's linearity violation gives us a complementary constraint to a new force in the intra-atomic (sub-keV to sub-MeV) range.

The rest of the paper is organized as follows.
In Sect.~\ref{SecNonlinear}, we revisit the breaking of King's linearity by the higher order effect of the field shift and by the new particle.
Afterwards, we present our numerical results of the violations of the linearity
in terms of the mass and the coupling of the new boson in Sect.~\ref{SecResult}.
We also compare the future sensitivity of King's linearity violation with other bounds.
Section~\ref{SecConc} is devoted to our conclusion.
\section{Non-linearity of the isotope shift}
\label{SecNonlinear}

We review King's linearity and discuss its violation in this section.
Two sources of the linearity violation are examined.
One is the next leading order (NLO) of the field shift, and the other is the light new mediator.
We call the isotope shift by the exchange of a new particle as the particle shift.

The isotope shift of a transition, which is represented by $\de\nu$, is described by the leading orders of the mass shift and the field shift, and the other contributions as
\begin{align}
 \de\nu = G \de\mu +F \de\avg{r^2} +X.
\label{EqIs}
\end{align}
The coefficients $G$ and $F$ represent the transition-dependent parts of the mass and the field shift, respectively.
The last term $X$ stands for the other contributions including the higher-order effects of the above shifts.
The reduced mass difference of the isotopes is denoted by $\de\mu$.
The transition-independent part of the field shift is given by the difference of the mean square charge radii $\de\avg{r^2}$ of the nuclei.
Since the precise charge distribution is not clear, to be compared to the reduced masses, we eliminate $\de\avg{r^2}$ with another transition.
Using the subscripts 1 and 2 for the different transitions, we obtain
\begin{align}
 \om_2 = \frac{F_2}{F_1} \om_1 + G_{21} + \frac{X_{21}}{\de\mu},
\label{EqModis}
\end{align}
where
\begin{align}
  \om_i =& \frac{\de\nu_i}{\de\mu},\\
	G_{21} =& G_2-\frac{F_2}{F_1} G_1,\\
	X_{21} =& X_2-\frac{F_2}{F_1} X_1.\label{Eq:X21}
\end{align}
We call $\om_i$ the modified isotope shift.
If the last term is independent of the mass numbers, the modified isotope shifts satisfy the linear relation, that is, King's linearity~\cite{RefKing}.

In the rest of this section, we firstly study the higher-order correction of the field shift.
Since it contributes $X$, the sensitivity to the new particle is limited by the size of its contribution.
Secondly, we discuss the particle shift.
The formulation is similar to the field shift.
Afterwards, the violation of the linearity is discussed in detail for both of the shifts.

\subsection{The field shift}

The nuclear charge distributions of the isotopes are slightly different from each other.
Their small modifications of the potentials are observed as the field shift in the isotope shift.

In the single electron approximation, assuming the spherical symmetry of the system, the field shift of a transition is written as
\begin{align}
 \text{FS} = \int dr\, r^2 \left( R_{N'}(r)^2 -R_N(r)^2 \right) \de V(r),
\end{align}
where $R_N(r)$ is the radial wave function of the state specified by a set of the quantum numbers represented as $N$. 
The modification of the potential $\de V(r)$ is defined as
\begin{align}
  \de V(\vec{r}) = -Z\al \int d\vec{r'}\, \frac{\de \rh(r')}{ \abs{\vec{r}-\vec{r'}} },
\end{align}
with the charge distribution difference $\de\rh (r)$ between two nuclei of charge $Z$. 
The spherical symmetry of $\de \rh$ is the origin of the symmetry of $\de V$.
The charge distribution $\rh$ is normalized as
\begin{align}
 4\pi \int dr\, r^2 \rh(r) =1.
\end{align}
Since the total charges of the isotopes are the same, the difference of the potential $\de V$ can be expressed as
\begin{align}
 \de V(r) =&
  -4\pi Z\al \left(
	 \int_0^r dr' \frac{r'^2}{r} \de\rh(r') +\int_r^\infty dr'\, r' \de\rh(r')
	\right) \\
 =& \label{Eq:dVint}
  -4\pi Z\al \int_r^\infty dr' \left(r'-\frac{r'^2}{r}\right)\de\rh(r').
\end{align}
In terms of the perturbation on the charge distribution difference, we do not have to take care the isotope dependence of the wave function at the leading order.

The charge distribution difference appears near the origin, and the potential difference does too.
The wave function close to the origin is modified from that given by the point charge, which is a good approximation away from the nucleus.
Therefore, we separately treat the wave functions in the two regions in calculating the field and the particle shifts.
The wave function near the nucleus is evaluated by the power series expansion at the origin, and then it is connected to the solution given by the point charge.
The connection point $r_c$ is also determined by the smooth connection condition.
A more explicit description of our treatment is shown in Appendix~\ref{AppOrigin}.

For simplicity, we discuss the contribution of a state to the field shift. 
We write the squared wave function inside the nucleus in a series expansion as
\begin{equation}
R_N(r)^2=r^{2l}\sum_{k=0}\xi_k^lr^k,
\end{equation}
where $l$ is the angular momentum.
The squared wave function outside the nucleus is denoted by $\Xi_l$.
The energy shift is given by the following integration:
\begin{align}
   \int_0^\infty dr\, r^2 R_N(r)^2 \de V(r)
 =&
   \int_0^{r_c} dr\, r^{2l+2} \sum_{k=0} \xi_k^l r^k \de V(r)
  +\int_{r_c}^\infty dr\, r^2 \Xi_l(r) \de V(r) \\ 
 =&
   \int_0^\infty dr\, r^{2l+2} \sum_{k=0} \xi^l_k r^k \de V(r)
	+\int_{r_c}^\infty dr\, r^2 \left( \Xi_l(r)-\sum_k \xi^l_k r^{k+2l} \right) \de V(r).
\label{EqFscalc}
\end{align}

With Eq.~\eqref{Eq:dVint} and the exchange of the integration order, the first term gives us the Seltzer moment expansion~\cite{PhRva.188.1916},
\begin{align}
   \int_0^\infty dr\, r^{l+2} \sum_{k=0} \xi^l_k r^k \de V(r)
 =&
   Z\al \sum_{k=0} \frac{\xi^l_k}{(k+2l+3)(k+2l+2)} \de\avg{r^{k+2l+2}},
\label{EqSeltzer}
\end{align}
where
\begin{align}
 \de\avg{r^a} =4\pi \int dr\, r^{a+2} \de \rho(r).
\end{align}
The explicit form of $\avg{r^a}$ with the Helm form factor \cite{Helm:1956zz} is shown in Appendix~\ref{AppHelm}.
For $l=0$, the leading term of the above expression is proportional to $\de\avg{r^2}$.
This term gives the field shift in Eq.~\eqref{EqIs}.
If the expansion of the wave function can be extended to the outside of the nucleus, the second term of Eq.~\eqref{EqFscalc} vanishes.

For the case of the Helm distribution, the ratio of the leading and the next leading-order field shift for $l=0$ is roughly estimated using the formulas shown in Appendices~\ref{AppOrigin} and \ref{AppHelm},
\begin{align}
 \frac{\text{NLO}}{\text{LO}} =&
   \frac{3}{10} \frac{\xi^0_2}{\xi^0_0}\frac{\de\avg{r^4}}{\de\avg{r^2}} \\
 =&
   \frac{3}{7}Z\al m_e r_N \left( 1 +6\frac{s^2}{r_N^2} +O\left( s^4/r_N^4 \right) \right) \\
 \sim &
   Z(4.10 +0.0255\, Z)\times 10^{-5}.
\end{align}

We have found the last approximate formula using $r_N = 0.519 +1.00 Z^{1/3} +0.103 Z^{2/3}$ fm, which is derived by fitting $r_N$ in Eq.~\eqref{Eq:rN2} as a function of $Z$ with the mass number $A$ being the standard atomic weight~\cite{RefNist}.
We have used the atoms of $Z\geq 10$ to find the above fit results.
For Ca$^+$ and Yb$^+$, the NLO-to-LO ratios are $8.69\times 10^{-4}$ and $4.15\times 10^{-3}$, respectively.
The estimate is consistent with Ref.~\cite{RefFs}.

\subsection{The particle shift}

Since the particle shift is sensitive to only the interaction of electron with neutron, we consider the following potential:
\begin{align}
 V_\text{PS}(r) = (-1)^{s+1} (A'-A) \frac{g_n g_e}{4\pi}\frac{e^{-mr}}{r},
\end{align}
where $g_n$ and $g_e$ are the coupling with neutron and electron, respectively.
The new force carrier is supposed to possess the spin $s$ and the mass $m$.
Then, in the single electron approximation, the particle shift is
\begin{align}\label{Eq:PS0}
 \text{PS} =&
   \int dr\,r^2 \left( R_{N'}(r)^2 -R_N(r)^2 \right) V_\text{PS} (r).
\end{align}
This is just given by the replacement of $\de V$ with the Yukawa potential in the field shift.

Following the discussion of the field shift, we consider the contribution to the particle shift by a state with the angular momentum $l$.
Omitting the couplings, it is written as
\begin{align}
 \int dr\,r^2 R_N(r)^2 \frac{e^{-mr}}{r}
 =&
   \int_0^{r_c} dr\,r^{2l+1} e^{-mr} \sum_{k=0} \xi^l_k r^k +\int_{r_c}^\infty dr\,r e^{-mr} \Xi_l(r) \\
 =&
   \sum_{k=0} \frac{\xi^l_k}{m^{k+2l+2}} (k+2l+1)!
	-\sum_{k=0} \frac{\xi^l_k}{m^{k+2l+2}} \Ga(k+2l+2,m r_c)
	+\int_{r_c}^\infty dr\,r e^{-mr} \Xi_l(r),
\label{EqPs}
\end{align}
where we use the incomplete gamma function, $\Ga(n,x) = \int_x^\infty dt\, t^{n-1} e^{-t}$.

For a given order of the Seltzer moment in Eq.~\eqref{EqSeltzer}, each term of the first summation in Eq.~\eqref{EqPs} is simultaneously eliminated with the corresponding field shift.

If the mass of the mediator is much larger than the inverse of the nuclear size, the particle shift is dominated by the leading term in the first summation of Eq.~\eqref{EqPs}.
Assuming that the transitions include $s$-states, the field shift coefficient $F$ is proportional to $\xi_0^0$ as shown in Eq.~\eqref{EqSeltzer}. 
The particle shift involved in $X$ is dominated by the contribution of $\xi_0^0$ in Eq.~\eqref{EqPs}. 
Thus $F\de\avg{r^2}+X$ is regarded as the product of $\xi_0^0$ and the transition-independent (but isotope-dependent) quantity.
Eliminating the latter with two transitions results in the linear relation in which the information of the particle shift is lost.
This is the reason why the violation of King's linearity is insensitive to the Higgs and $Z$ bosons, as discussed in Ref.~\cite{1601.05087}.
For a light mediator, the integration in Eq.~\eqref{EqPs} gives us the leading contribution.
If the mass of the mediator is small enough to cover the whole wave function, this contribution becomes independent of the mass.
\subsection{Non-linearity}

So far, we have obtained some additional contributions to the isotope shift.
Including the terms discussed in this section, Eq.~\eqref{EqIs} is modified as
\begin{align}
 \de\nu = G\de\mu +F\de\avg{r^2} +\tilde{F} + (A'-A) H.
 \label{EqModis2}
\end{align}
The higher-order field shift is described by $\tilde{F}$, and its leading contribution is proportional to $\de\avg{r^4}$. 
The last term including $H$ is the contribution of the particle shift.
These terms have appeared as $X$ in Eq.~\eqref{EqIs}, then, in general, they violate King's linearity.

In Eq.~\eqref{EqModis}, the values $F_{1,2}$ are proportional to the differences of the squared wave functions at the origin (see Eq.~\eqref{EqSeltzer}).
Omitting the electron spin degree of freedom as the numerical analysis in Sect.~\ref{SecResult}, three or four distinct electronic orbital states are involved in the two transitions. 
Here we consider the case that one of the orbital states is an $s$-state and the others are non-$s$-states.
We observe the following features of the non-linearity:
\begin{itemize}
 \item If both of the transitions include the $s$-state, its contributions are canceled each other.
 \item If only one of the transitions includes the $s$-state, its contribution is suppressed.
\end{itemize}
In the former case, we obtain $X_2/X_1= F_2/F_1$ as far as the $s$-state contributions in $X_{1,2}$ are considered, so that the $s$-state contribution to $X_{21}$ in Eq.~\eqref{Eq:X21} vanishes.
In the latter case, assuming that the transition 1 includes the $s$-state, namely, $F_1\neq 0$ and $F_2=0$, then $X_{21}=X_2$.
That is, the dominant contribution due to the $s$-state is suppressed.
In both of the cases, the non-linearity is induced by higher angular momentum states.
We shall use the experimental results of these situations in the numerical analysis later.
To obtain effects of the non-linearity by the $s$-state, the transitions need to include at least two distinct $s$-states.

The non-linearity given by the particle shift follows the same features.
The contributions of the $s$-states disappear unless the relevant transitions include two or more distinct $s$-states.
Then only the states with the higher angler momenta contribute to the non-linearity.
The sensitivity to the new force carrier is scaled with an inverse power of its mass if it becomes much heavier than the typical scale of the wave function.
Since the leading contribution is not given by the $s$-state in the above situation, the scaling is $1/m^4$ or worse.

Finally, we mention the extension of King's linearity.
The higher-order corrections, such as the NLO field shift above, prevent us from improving the experimental bounds of the light mediator search.
However, it is possible to eliminate the effect of the NLO field shift and obtain an extended linear relation if we use additional data of isotope shifts with another transition.
The general structure of the linearity is discussed in Appendix~\ref{AppElim}.

If we eliminate the higher-order field shift, the corresponding contributions of the particle shift are simultaneously removed.
This means that the above scaling in the heavier mediator region becomes worse.
Including distinct $s$-states helps us to improve the sensitivity in a region between the atomic and the nuclear scales.
\section{Numerical analysis}
\label{SecResult}

We show the current status and the future prospect of the new weakly interacting light mediator search with the isotope shift.

The particle shift is measured as the deviation from King's linearity with two transitions of an element.
Hence, in addition to the experimental precision, the choices of elements and transitions are also important to measure the effect of the new force.
In the following numerical analysis, we compare the calcium ion Ca$^+$ and the ytterbium ion Yb$^+$. 
As shown below, some isotope shifts of Ca$^+$ are precisely measured at present and the data satisfy King's linearity within the errors.
The relative errors in its isotope shifts are better than $10^{-4}$.
Experiments on Yb$^+$ give us similar bounds on the particle shift though the errors are about one order of magnitude worse than those of Ca$^+$.
We investigate the future prospect of experimental bounds, simply reducing the experimental error to 1 Hz as an illustration.

The precise wave functions of the states in the transitions are quite difficult to obtain for heavier elements because it is a many-body problem.
We calculate them as the wave function of an electron in an effective potential given by other electrons and the nucleus.
The effective potential is estimated by the Thomas-Fermi model, which is a semi-classical approximation of the electrons around the nucleus; see e.g.~\cite{March}.
Our analysis is done in the non-relativistic limit.
In this case, the wave function does not discriminate the spin dependence.
The electronic states are characterized by a pair of quantum numbers $(n,l)$.
This approximation is good for $s$- and $p$-states as shown in Appendix~\ref{AppTF} and Ref.~\cite{PhRva99.510}.
The same data set of Ca$^+$ isotope shifts is studied in Ref.~\cite{1704.05068} using different wave functions from ours. 
Both results reasonably agree in the sufficiently small mass region.
\subsection{Experimental data of isotope shifts}
The experimental data of the isotope shifts used in our analysis are summarized in Table~\ref{TabIs}. The masses of isotopes for the calculation of the modified isotope shifts are given in Ref.~\cite{RefNist}.

\begin{table}[tb]
 \centering
 \begin{tabular}{c|rr}
  Ca$^+$ & 397 nm & 866 nm \\ \hline
	    42 & 0.425 706(94) & -2.349 974(90) \\
	    44 & 0.849 534(74) & -4.498 883(80) \\
	    48 & 1.705 389(60) & -8.297 769(81) 
 \end{tabular}
 \qquad
 \begin{tabular}{c|rr}
   Yb$^+$ & 369 nm & 935 nm \\ \hline
	    170 & -1.6233(8)\hspace{.5em} & -3.464(10) \\
	    174 &  1.2753(7)\hspace{.5em} &  2.612(3)\hspace{.5em} \\
	    176 &  2.4928(10) &  5.074(19)    
 \end{tabular}
 \caption{
  Isotope shifts of Ca$^+$ and Yb$^+$ in the unit of GHz.
	The shifts of Ca$^+$ are given by Ref.~\cite{PRLta.115.053003}, and those of Yb$^+$ are given by Refs.~\cite{PhRva.A49.3351} (369 nm) and \cite{RefSugiyama} (935 nm).
	The shifts are measured from the mass number of 40 for Ca$^+$ and 172 for Yb$^+$.
 }
 \label{TabIs}
\end{table}

For Ca$^+$, we use the transitions of 397 nm ($^2S_{1/2}\to {}^2P_{1/2}$) and 866 nm ($^2D_{3/2}\to {}^2P_{1/2}$) reported in Ref.~\cite{PRLta.115.053003}.
In our calculations, the first one is considered as the transition of $4s\to4p$ and the second one is the $3d\to4p$ transition.

\begin{figure}[tb]
\centering
 \includegraphics[scale=.5,clip]{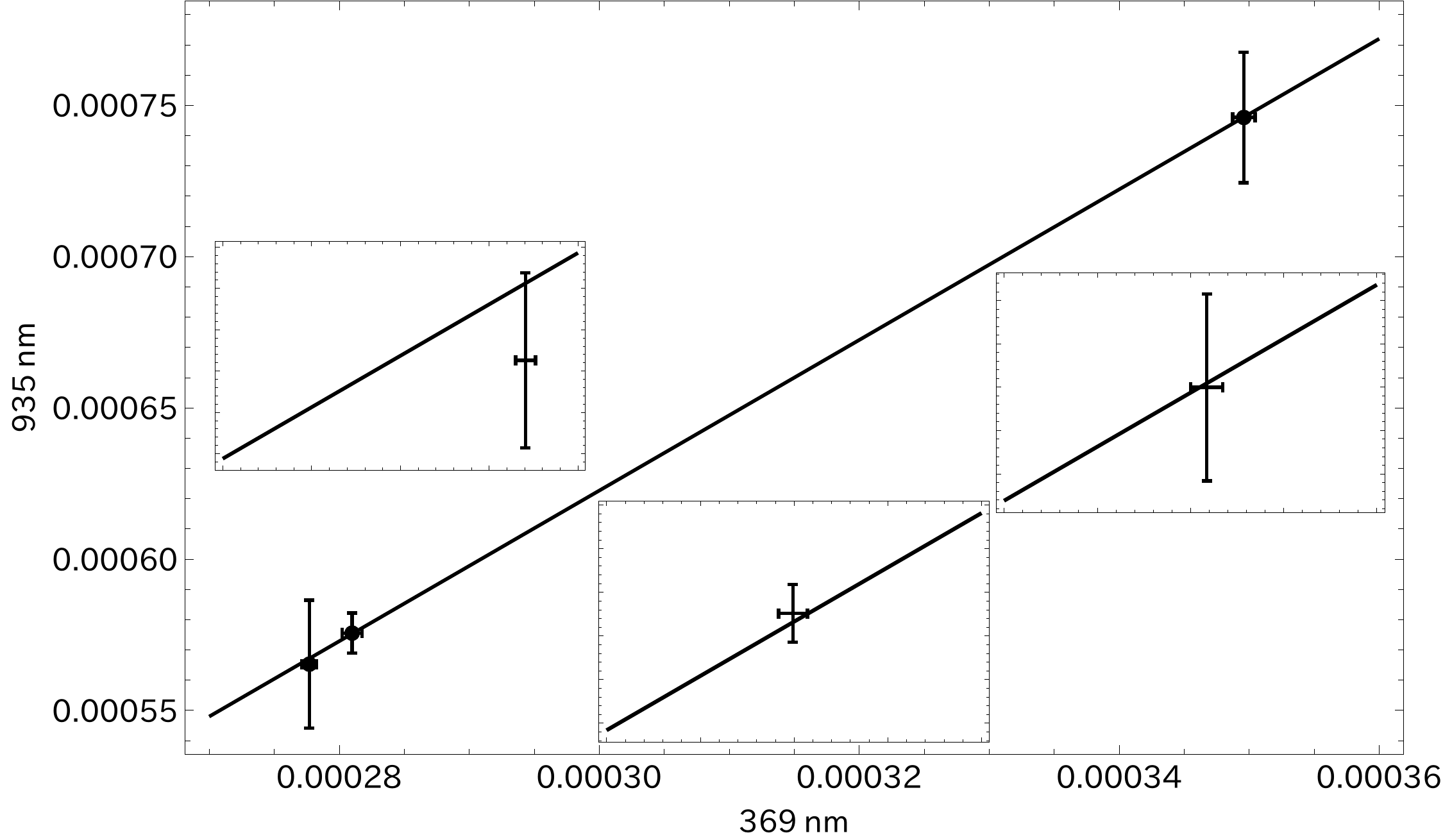}
 \caption{
 King's plot of Yb$^+$. The modified isotope shifts are in the natural unit.
 In the main figure, the error bars are 10$\si$, while they are 1$\sigma$ in the insets.
 The width and the height of the insets are $2\times 10^{-6}$ and $5\times 10^{-6}$, respectively. The isotope pairs of the plotted points are (176,172) (174,172) and (172,170) from left to right.
 }
 \label{FigKing}
\end{figure}

For Yb$^+$, the transitions of 369 nm ($^2S_{1/2}\to {}^2P_{1/2}$) and 935 nm ($^2D_{3/2}\to {}^3D[3/2]_{1/2}$) are studied.
They are given in Refs.~\cite{PhRva.A49.3351} and \cite{RefSugiyama}, respectively.
These transitions, respectively, correspond to the $6s\to6p$ transition and the $4f\to6s$ transition in our study.
For these transitions, King's plot is shown in Fig.~\ref{FigKing}.

Both of transitions include only one $s$-state, $4s$ for Ca$^+$ and $6s$ for Yb$^+$.
As discussed in the previous section, the $s$-states do not contribute to the non-linearity of King's plot in these cases.
Besides, the field shift of the higher Seltzer moments rapidly become small.
Therefore, the $p$-states numerically dominate the field shift non-linearity.

The constraints to the non-linearity are calculated by the usual $\ch^2$ as described in Appendix~\ref{AppStat}.
The bound to $g_n g_e$ depends on its sign.
The weaker bound is employed as the one for its absolute value.

\subsection{Current experimental bounds and future prospects}

\begin{figure}[tb]
\centering
 \includegraphics[scale=.6,clip]{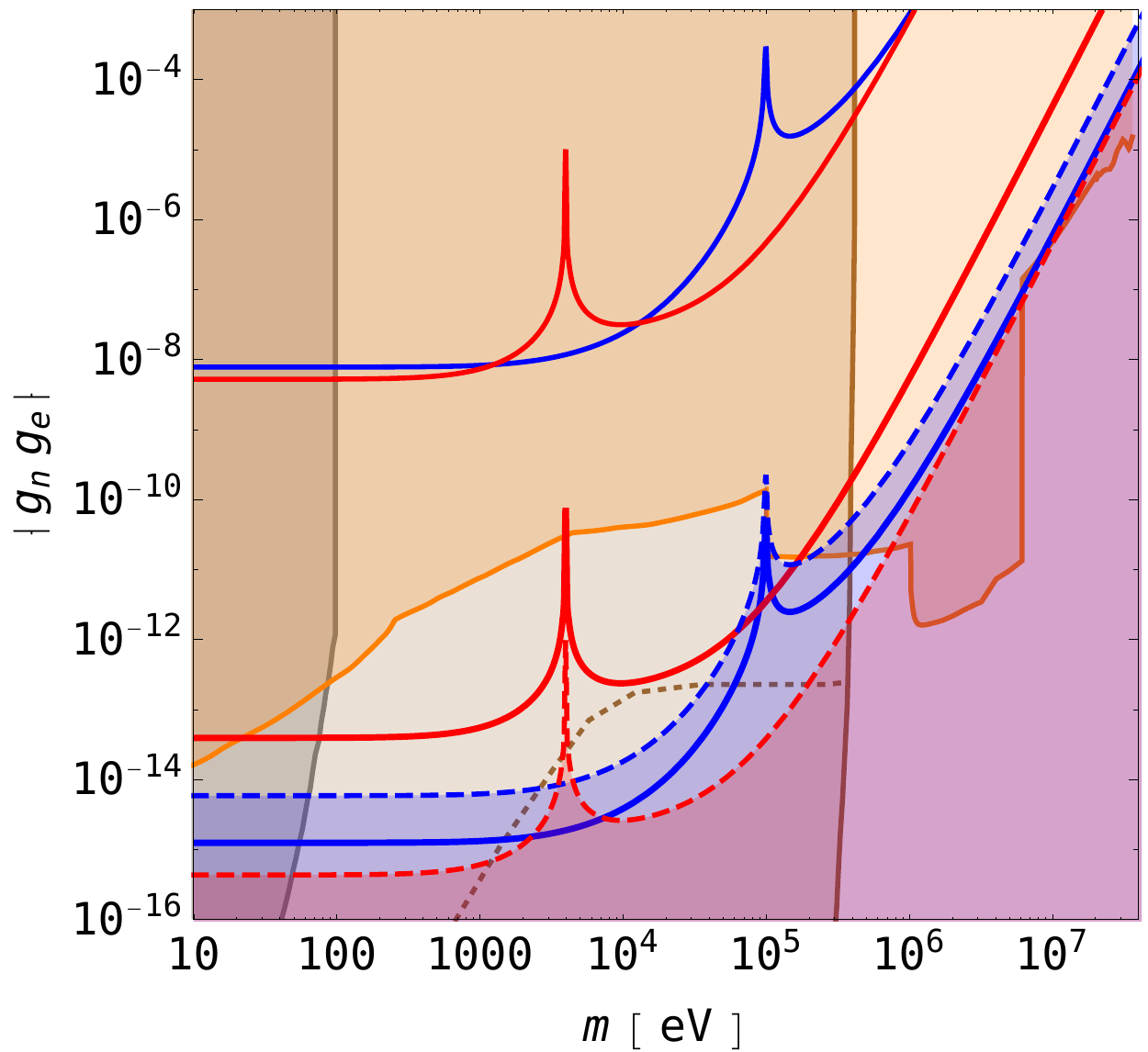}
\qquad
 \includegraphics[scale=.6,clip]{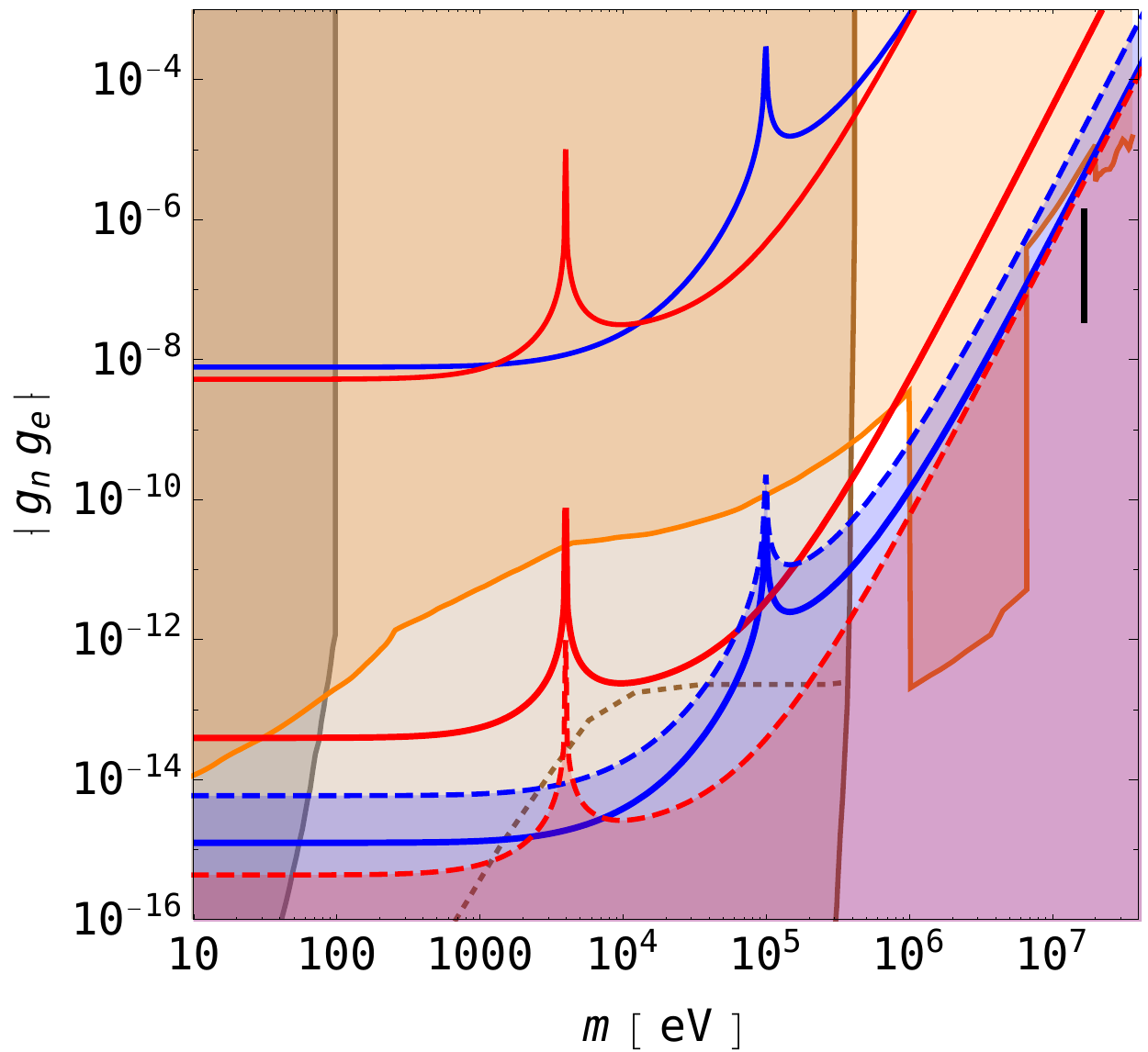}
 \caption{
  The isotope shift and other experimental constraints on the mass and the coupling of the light force carrier.
	The left panel is for the scalar mediator and the right one is for the vector mediator.
	The red/blue lines and region are used for the bounds with the isotope shifts by Ca$^+$/Yb$^+$.
	The upper solid lines are the current experimental bounds, and the lower ones are the future prospects with the error of 1 Hz.
	The non-linearities by the NLO field shifts appear in the shaded regions below the dashed lines.
	The upper shaded orange regions indicate the constraints obtained with the bounds on the couplings of electron and neutron by low energy experiments, see the main text for the details.
	The shaded brown regions below about 500 keV is constrained by the stellar cooling bounds given by Ref.~\cite{An:2013yfc}.
	As described in Ref.~\cite{Redondo:2008aa}, the stellar bounds have uncertainty above the brown dotted line.
	The shaded gray regions below about 100 eV are restricted by the fifth force experiment~\cite{Ederth:2000zz,Fischbach:2001ry}.
The black line in the right panel stands for the region indicated by the Atomki anomaly~\cite{Krasznahorkay:2015iga,Feng:2016jff}.
 }
 \label{FigIs}
\end{figure}

The current bounds and the future prospects of King's linearity violation are shown in Fig.~\ref{FigIs}.
The lines given by the isotope shifts are the same in both panels.
The left and the right panels show other experimental constraints on the scalar and the vector mediators, respectively.

The integrand of the particle shift in Eq.~\eqref{Eq:PS0} includes the difference of the squared wave functions.
If the mediator is massless, both of the states contribute the integral.
Since the integrand flips the sign at a point, the integral disappears around the corresponding mass scale.
This cancellation explains the peak structures in Fig.~\ref{FigIs}.
The position of cancellation depends on the combination of the wave functions.
This means that testing the linearity with various atoms and transitions is important not only to check each against the other but also to exclude the cancellation points.

For the current bounds by the isotope shifts, Ca$^+$ and Yb$^+$ give us similar bounds, although the experimental errors of Yb$^+$ is about one order of magnitude worse than those of Ca$^+$.
Accordingly, the sensitivity of Yb$^+$ is about one order of magnitude better than that of Ca$^+$ in the prospected bounds with the error of 1 Hz.

The field-shift non-linearity of Ca$^+$ and Yb$^+$ appear at $1.1\times 10^{-2}$ Hz and 4.7 Hz, respectively.
These frequencies are interpreted in terms of the mass and the coupling of the light mediator as indicated by the dashed lines in Fig.~\ref{FigIs}.
Once the experimental sensitivity reaches the line, the bound on the particle shift is not improved without the further subtraction of the field-shift non-linearity as discussed in Appendix~\ref{AppElim}.
The Yb$^+$ line of the expected sensitivity lies below the field-shift non-linearity indicated by the dashed line, while the Ca$^+$ line does not.
If the precision of the Ca$^+$ measurement is improved, it may cover the region of smaller coupling.
However, in this case, the non-linearity of the NLO mass shifts, which are more significant for lighter elements, also ought to be taken into account.

The experimental bounds are obtained as the allowed size of the non-linearity.
They depend on the sign of the coupling with the light new mediator.
Hence, the constraints to the mediator are changed at the peaks since the sign of the particle shift is changed there.
As already mentioned, we have adopted the sign giving the weaker constraint to calculate each of the current experimental bounds. Thus, the current bounds shown in Fig.~\ref{FigIs} are the conservative ones.

Now, we turn to other constraints by light particle searches in the same mass region.

We consider the terrestrial bounds which are obtained as the product of the individual bounds on the couplings with electron and neutron.
The constraint of the electron coupling is mostly given by the electron $g-2$~\cite{Hanneke:2008tm}.
The region above 20 MeV is bounded by Babar~\cite{Lees:2014xha}.
The beam dump experiments strongly constrain the region from 100 keV and 1 MeV to about 10 MeV for the scalar and the vector mediators, respectively~\cite{Liu:2016qwd,Andreas:2012mt}.
As summarized in Ref.~\cite{PhRva.D96.015011}, the experimental bounds on the neutron coupling is given by the combination of several low energy neutron experiments~\cite{PRLta.68.1472,Pokotilovski:2006up,Nesvizhevsky:2007by}.
The obtained bounds are much stronger than the current bounds by the isotope-shift non-linearity.
However, the future prospects of the sensitivities are better than the above terrestrial bound in the region between about 10 eV and 1 MeV.

The mass region less than about 100 eV is covered by the fifth force search~\cite{Ederth:2000zz,Fischbach:2001ry}.
Since this bound is very strong, the mediator lighter than about 100 eV is almost excluded for the entire coupling region plotted in Fig.~\ref{FigIs}.

The stellar cooling bounds are also very strong in the region lighter than about 1 MeV~\cite{An:2013yfc}.
However, if the coupling becomes large, the light new boson cannot take the energy to the outside.
It is not sure that the stellar cooling observation can constrain the boson for sufficiently strong couplings. 
The brown dotted line in Fig.~\ref{FigIs} indicates this limitation given by Ref.~\cite{Redondo:2008aa}. 
We consider the isotope-shift non-linearity as well as other terrestrial experiments to be useful since it provides us constraints independent of the stellar dynamics.

The mass and the coupling of the light vector boson suggested by the Atomki $^8$Be experiment~\cite{Krasznahorkay:2015iga,Feng:2016jff} are also shown in the right panel of Fig.~\ref{FigIs}.
The result in Ref.~\cite{1704.05068} is that the region can be excluded by the future Yb$^+$ measurement.
However, our result indicates that the sensitivity cannot reach the coupling of the vector.
Even worse, the field-shift non-linearity prevents us from probing the parameter region.
We note that the present work and Ref.~\cite{1704.05068} employ different transitions of Yb$^+$, and our result is derived from the existing data of the Yb$^+$ isotope shifts.

\section{Conclusion}
\label{SecConc}

The isotope shifts can be precisely measured with the technique developed in the atomic clock experiments.
In the measurements, the weakly interacting light force carrier can be tested as the violation of King's linearity.

We have investigated two sources of the non-linearity.
One is the higher-order effect of the field shift and the other is the above effect of the light mediator, the particle shift.

Because of the larger particle shift of heavier elements, the sensitivity to the light mediator of Yb$^+$ is about one order of magnitude better than that of Ca$^+$.
However, Yb$^+$ also possesses the sizable field-shift non-linearity.
In the future prospects with the 1 Hz error, the sensitivity of Yb$^+$ reaches the non-linearity given by the field shift.
Then its sensitivity to the light mediator is not improved without the further development of the formulation of the linearity.
On the other hand, the field-shift non-linearity of Ca$^+$ is about one order of magnitude smaller than that of Yb$^+$ in terms of the coupling of the light mediator.

The future prospect gives us the bounds better than the other low energy experiments in the region from 100 eV to 1 MeV.
This region is also bounded by the observation of the stellar cooling.
The terrestrial experiments including the isotope-shift non-linearity tell us the complementary information.

In our study, the scaling of the sensitivity is analytically clarified as $1/m^4$ in the region where the force carrier is heavy.
In the numerical analysis, this behavior is indeed observed if the boson is heavier than about 100 keV.
As a result, the expected exclusion lines with the error of 1 Hz cannot reach the parameter region of the Atomki anomaly.
Furthermore, the favored region cannot be proved because of the field shift non-linearity even if the experimental precision is improved.

Our analysis uses the non-relativistic limit, the mean field approximation with the Thomas-Fermi potential, and the assumption of the Helm distribution.
It is required to improve the theoretical methods for more precise and extensive studies of the isotope-shift non-linearity including other atoms and transitions.

\section*{Acknowledgments}
The authors thank K.~Sugiyama for informing them as regards his data of the $\text{Yb}^+$ isotope shifts.
M.T. also thanks N.~Sasao for useful discussions.
The work of M.T. is supported in part by JSPS KAKENHI Grant numbers JP 25400257, 15H02093, 16H03993, 17H02895, and 17H05405.
The work of Y.Y. is supported in part by the National Research Foundation of Korea Grant no. 2016R1A2B2016112.
\appendix
\section{Wave function inside the nucleus}
\label{AppOrigin}

The Schr\"odinger equation of the radial direction is
\begin{align}
 \left( \frac{d^2}{dr^2} -\frac{l(l+1)}{r^2} +2m_e(E-V(r))  \right) r R_N(r) =0.
\end{align}
Since we consider a bound state, the potential $V(r)$ and the energy $E$ are negative in this equation.

We are interested in the wave function near the origin.
The potential is expanded as
\begin{align}\label{Eq:SeriesV}
 V(r) = \sum_{i=0} v_i r^i,
\end{align}
where $v_0<0$, and $v_1=0$ for a nuclear charge distribution without a cups at the origin such as the Helm distribution described in Appendix~\ref{AppHelm}.
It is supposed that the wave function is also given as the following series:
\begin{align}
 R_N(r) = \sum_{i=0} \ch^l_i r^{i+l}.
\end{align}
Then the Schr\"odinger equation is expressed as
\begin{align}
 0= 2(l+1) \ch^l_1 +\sum_{i=0} r^{i+1} \left(
  (2l+i+3)(i+2) \ch^l_{i+2} +2m_e( (E-v_0) \ch^l_i -\sum_{j=0}^{i-2} v_{i-j} \ch^l_j )
 \right).
\end{align}
The first term leads to $\ch^l_1 = 0$.
The entity within the large parentheses gives us the recurrence relation to determine the coefficients $\ch^l_i$.
The first equation of the above relation with $i=0$ is
\begin{align}
 0 =& (2l+3) \ch^l_2 +m_e(E-v_0) \ch^l_0.
\end{align}
Using $E/v_0\ll 1$, we obtain
\begin{align}
 \frac{\ch^l_2}{\ch^l_0} &= \frac{m_e v_0}{2l+3}.
\end{align}
The rest of the coefficients are recursively obtained.
We use the wave function of $O(r^2)$.
The squared wave function inside the nucleus is 
\begin{align}
 R_N (r)^2
 =&
   r^{2l} \ch^l_0{}^2 \left( 1 +\frac{2 m_e v_0}{2l +3} r^2 +\cdots \right).
\end{align}
This wave function is smoothly connected to the wave function with the point charge, i.e., the solution outside the nucleus.

The coefficient $\ch^l_0$ and the connection point $r_c$ are both obtained by the condition to connect.
It is convenient to express the wave function with the point charge as $r^l(a_0+a_1 r)$ near the nucleus.
Then the above quantities are analytically calculated as
\begin{align}
 \ch^l_0 =&
   \frac{1}{2} \left( a_0 - \frac{\sqrt{m_e v_0 \left(a_0^2 m_e v_0 +(2l+3) a_1^2\right)}}{m_e v_0} \right)
	\sim 
   a_0 +\frac{(2l+3)a_1^2}{4a_0 m_e v_0} ,\\
 r_c =&
  -\frac{a_0 m_e v_0 +\sqrt{m_e v_0 \left(a_0^2 m_e v_0 +(2l+3) a_1^2\right)}}{a_1 m_e v_0}
	\sim
	 \frac{(2l+3) a_1}{2a_0 m_e v_0}.
\end{align}

Strictly speaking, the wave function inside the nucleus varies from one isotope to another because $v_0$ depends on $r_N$ as shown in Eq.~\eqref{Equationnnnn}. 
Since this is a higher-order effect, we use the fixed $r_N$ of $A=44$ for all the Ca isotopes and $A=173$ for the Yb ones.

\section{The nuclear charge density and the potential with the Helm distribution}
\label{AppHelm}

The Helm distribution of the nuclear charge~\cite{Helm:1956zz} is defined as
\begin{align}
    \tilde{\rh}_\text{Helm} (q)
 &= 3 \frac{\sin(q r_N) -qr_N \cos(qr_N)}{(qr_N)^3} e^{-q^2 s^2 /2}.
\end{align}
In our numerical analysis, we use the following parameters given by Ref.~\cite{Lewin:1995rx}:
\begin{align}
 r_N^2 &= c^2 +\frac{7}{3}\pi^2 a^2 -5s^2,\label{Eq:rN2}\\
		 a &\sim 0.52 \text{ fm},\\
     c &\sim 1.23 A^{1/3} -0.60 \text{ fm},\\
		 s &\sim 0.9 \text{ fm}.
\end{align}
This distribution is given by smearing of the uniform density distribution as shown below.
The limit of $s\to 0$ corresponds to the uniform distribution with the radius of $r_N$.

The spatial density is obtained:
\begin{align}
    \rh_\text{Helm} (r)
 &= \int \frac{d\vec{q}}{(2\pi)^3}\, \tilde{\rh}_\text{Helm} (q) e^{i \vec{q}\cdot\vec{r}},\n
 &= \frac{3}{8\pi^2 r_N^3 r} \left(
    -2s\sqrt{2\pi} \sinh \left( \frac{r r_N}{s^2} \right) e^{-\frac{r^2 +r_N^2}{2s^2}}
		+\pi r \left(
		  \text{Erf} \left( \frac{r+r_N}{\sqrt{2}s} \right) -\text{Erf} \left( \frac{r-r_N}{\sqrt{2}s} \right)
		 \right) 
	 \right),
\end{align}
where the error function Erf$(x)$ is defined as 
\begin{align}
  \text{Erf}(x) = \frac{2}{\sqrt{\pi}} \int_0^x dt e^{-t^2}.
\end{align}
The distribution is also expressed as
\begin{align}
    \rh_\text{Helm}(r)
 &= \int d^3r'\, \frac{3}{4\pi r_N^3} \th(r_N-r) \rh_g(\vec{r}-\vec{r'}),
\end{align}
where $\th$ is the step function and the smearing function $\rh_g$ is given by the Gaussian distribution as
\begin{align}
  \rh_g (\vec{r}) = \frac{e^{-r^2/2s^2}}{(2\pi s^2)^{3/2}}.
\end{align}
We can check the whole space integration of this density indeed becomes unity, namely,
\begin{align}
  4\pi \int_0^\infty dr\, r^2 \rh_\text{Helm} (r) = 1.
\end{align}

The electrostatic potential induced by the above density is
\begin{align}
 V(r)
 =&
   -Z\al\int_0^\infty d\vec{r'} \frac{\rh_\text{Helm} (\vec{r'})}{\abs{ \vec{r}-\vec{r'} }} \\
 =&
	 -\frac{Z\al}{4\pi r_N^3 r} \biggl(
      \sqrt{2\pi} s\left(
			   (r^2 +r_N r -2r_N^2 +2s^2) e^{-\frac{(r+r_N)^2}{2s^2}}
				-(r^2 -r_N r -2r_N^2 +2s^2) e^{-\frac{(r-r_N)^2}{2s^2}}
			\right) \n &
	  -\pi \left(
		  (r^3 -3(r_N+s)(r_N-s)r -2r_N^3) \text{Erf}\left(\frac{r+r_N}{\sqrt{2}s}\right)
		 -(r^3 -3(r_N+s)(r_N-s)r +2r_N^3) \text{Erf}\left(\frac{r-r_N}{\sqrt{2}s}\right)
		\right)
	 \biggr).
\end{align}

This function is even with respect to $r$.
The potential at the origin is
\begin{align}
  v_0 =& -3\frac{Z\al}{2r_N} \left(
	   \left(1-\frac{s^2}{r_N^2}\right) \text{Erf}\left( \frac{r_N}{\sqrt{2}s} \right)
		+\frac{s}{r_N}\sqrt{\frac{2}{\pi}} e^{-\frac{r_N^2}{2s^2}}
   \right).
 \label{Equationnnnn}
\end{align}

In the Seltzer moment expansion, we need the mean values of $r^n$,
\begin{align}
  \avg{r^n}
 =&
   4\pi \int dr\, \rh(r)r^{2+n}\\
 =& 
   \frac{2(\sqrt{2}s)^n}{\sqrt{\pi}} \Ga\pfrac{n+3}{2}
	 {}_1F_1\left(-\frac{n}{2},\frac{5}{2},-\frac{r_N^2}{2s^2} \right),
\end{align}
where we introduce the confluent hyper geometric function of the first kind, 
\begin{align}
  {}_1F_1(a,b,z) = \frac{\Ga(b)}{\Ga(b-a)\Ga(a)}\int_0^1 dt\, e^{zt} t^{a-1} (1-t)^{b-a-1}.
\end{align}
The relevant terms in our calculations are
\begin{align}
  \Avg{\frac{1}{r}} =& -\frac{v_0}{Z\al},\\
  \avg{r^2} =& \frac{3}{5} (r_N^2 +5s^2),\\
  \avg{r^4} =& \frac{3}{7} (r_N^4 +14r_N^2 s^2 +35s^4).
\end{align}
\section{Generalization of the linearity}
\label{AppElim}

Eliminating the difference of the mean squared radii with two distinct transitions, King's linearity is obtained.
It turns out that the higher-order field shift and even the particle shift can be similarly eliminated with additional transitions.
We formulate this procedure and derive the generalization of the linearity as follows.

For simplicity, we discuss it with the higher-order field shifts at first.
The isotope shift is written as
\begin{align}
 \de\nu_i = G_i \de\mu +\sum_k F_i^{(k)} \de\avg{r^{2+k}} +X_i,
\end{align}
where the terms including $F_i^{(k)}$ are the higher-order field shifts in terms of the Seltzer moment.
The last term $X_i$ stands for the other non-linearity.
If we have a sufficient number of transition data, the above expression can be rewritten as
\begin{align}
  \de\vec{\nu} -\vec{X}
 = 
  \begin{pmatrix}
    \vdots &    \vdots & \vdots    &        \\
	 	   G_i & F_i^{(0)} & F_i^{(2)} & \cdots \\
    \vdots &    \vdots & \vdots    &        
	\end{pmatrix}
	\begin{pmatrix}
    \de\mu \\ \de\avg{r^2} \\ \de\avg{r^4} \\ \vdots
	\end{pmatrix}.\label{Eq:ISs}
\end{align}
For convenience, we write the matrix in the right hand side as $T$.
If $T$ has an inverse, we multiply it to the above equation.
The difference of the reduced masses $\de\mu$ is precisely measured, while the mean radii are not.

Multiplying $T^{-1}$ to the above equation, the first element is 
\begin{align}
 \sum_i \left( T^{-1} \right)_{1i} \left( \de\vec{\nu} -\vec{X} \right)_i = 
 \delta\mu.
\end{align}
Dividing this equation by $\de\mu$, we obtain the generalized linear relation which is free of the field-shift non-linearity.
If $X_i=0$ for any $i$, the data of the modified isotope shifts with $n$ different transitions are on an $n-1$ dimensional plane.

The above procedure is easily extended to the other non-linearity.
Firstly, the given term is separated into the wave-function-dependent part and -independent part.
This way of separation is not unique, however, the result is independent of this detail.
Secondly, the wave-function-dependent part is embedded in the above matrix $T$, and then we multiply the inverse.
Finally, the precisely measured element, like $\de\mu$ above, gives us the linear relation we want.
This means that we may simultaneously obtain several linear relations.

For example, we consider the elimination of the non-linearity by the particle shift.
In order to do so, the factor independent of the wave function $(A'-A)$ is appended to the above vector, and the other part is embedded in $T$.
Then a new linear relation is derived in the same manner as above.
Since the factor $(A'-A)$ is also determined with a high precision, the additional linear relation is found from the different element.
Even if King's linearity violation is observed, these generalized linear relations are preserved as long as the non-linearity originates from the particle shift.

\section{The Thomas-Fermi potential}
\label{AppTF}

The Thomas-Fermi model is a semi-classical approximation of the electrons around the nucleus; see e.g.~\cite{March}.
The electrons are regarded as the free fermion gas of zero temperature.
Two electrons occupy the phase space of $(2\pi \hbar)^3$ from the bottom.
The self-consistent potential is represented by the universal function so called the Thomas-Fermi function.

This function is the solution of the following non-linear differential equation:
\begin{align}
  \frac{d^2 \ch}{dx^2} =\sqrt{\frac{\ch^3}{x}}.
\end{align}
One boundary condition is given at the origin as $\ch(0) = 1$.
For positive ions, the other boundary condition is imposed to satisfy
\begin{align}
 x_0 \ch'(x_0) =-\frac{n}{Z},
\end{align}
where $\ch(x)$ vanishes at $x_0$, and $n$ is the positive charge of the ion.
Since we consider an electron in a singly charged positive ion, the mean potential is given by the other electrons and the nucleus, namely, $n=2$.
The resulting potential energy is
\begin{align}
 V_\text{TF}(r) = 
 \begin{cases}
   -\frac{Z\al}{r} \ch(x) -n\frac{\al}{r_0} & 0<x<x_0 \\
	 -n\frac{\al}{r} & x_0 < x 
 \end{cases},
\end{align}
where $x =4\sqrt[3]{2 Z / 9\pi^2}\, m_e \al r$.
The boundary $x_0$ is approximately given by $x_0 = -8.964 + 7.341 Z^{1/3}$.
The wavelengths of the relevant transition evaluated in the Thomas-Fermi potential are shown in Table~\ref{TabEigen} as well as the corresponding experimental data.
\begin{table}[tb]
 \centering
 \begin{tabular}{c|rr}
    Ca$^+$ &  EX & TF \\ \hline
	4p$\to$4s & 397 & 475 \\
	4p$\to$3d & 866 & -1610 
 \end{tabular}
   \qquad
 \begin{tabular}{c|rr}
    Yb$^+$ &  EX &  TF \\ \hline
	6p$\to$6s & 369 & 380 \\
	6s$\to$4f & 935 & 48.6  
 \end{tabular}
\caption{
 The transition wavelengths for Ca$^+$ and Yb$^+$ in the unit of nm.
 The columns of EX stand for the experimental values, and those of TF are given by the Thomas-Fermi potential.
}
\label{TabEigen}
\end{table}

\section{Statistics}
\label{AppStat}

We use the following formulae in the numerical analysis.
The data of the modified isotope shifts of two transitions are denoted by $x_a$ and $y_a$, and their standard deviations are $\si_{xa}$ and $\si_{ya}$, respectively.
The subscript $a$ indicates an isotope pair.
The term violating the linearity is represented by $\ep s_a$.
The parameter $\ep$ stands for the wave function independent part, e.g., the coupling of the particle shift.
The $\ch^2$ of the fit function can be written as
\begin{align}
 \ch^2 = \sum_a \left( 
   \frac{(x_a -\hat{x}_a)^2}{\si_{xa}^2} +\frac{(y_a -f\hat{x}_a -g -\ep s_a)^2}{\si_{ya}^2}
 \right).
\end{align}
The parameter $\hat{x}_a$ stands for the point on the fit line to evaluate the $
\ch^2$.
The other parameters $f$, $g$ and $\ep$ are the fitting variables.
These parameters are chosen to minimize the $\ch^2$.
The minimization condition for $\hat{x}_a$ is given by 
\begin{align}
 \frac{\del \ch^2}{\del \hat{x}_a} =&
  -2\frac{x_a -\hat{x}_a}{\si_{xa}^2} -2f\frac{y_a -f\hat{x}_a -g -\ep s_a}{\si_{ya}^2} =0,
\end{align}
then,
\begin{align}
 \hat{x}_a =\frac{\si_{ya}^2 x_a +f\si_{xa}^2(y_a-g-\ep s_a)}{\si_{ya}^2+f^2\si_{xa}^2}.
\end{align}
Substituting it to the original $\ch^2$, we obtain
\begin{align}
 \ch^2 = \sum_a \frac{(y_a-fx_a-g-\ep s_a)^2}{\si_{ya}^2+f^2\si_{xa}^2}.
 \label{EqChi2}
\end{align}

The stability conditions for the rest of the variables are
\begin{align}
 \frac{\del \ch^2}{\del f} =&
   -2\sum_a \frac{(y_a -fx_a -g-\ep s_a)(f\si_{xa}^2(y_a-g-\ep s_a)+\si_{ya}^2 x_a)}{(\si_{ya}^2 +f^2\si_{xa}^2)^2} =0,\\
 \frac{\del \ch^2}{\del g} =&
   -2\sum_a \frac{y_a-fx_a-g-\ep s_a}{\si_{ya}^2 +f^2 \si_{xa}^2} =0,\\
 \frac{\del \ch^2}{\del \ep} =&
   -2\sum_a s_a \frac{y_a-fx_a-g-\ep s_a}{\si_{ya}^2 +f^2 \si_{xa}^2} =0.
\end{align}
These conditions show us the $\ch^2$ minimum, which is zero in the present work because of the additional parameter $\ep$.
We calculate the $\ch^2$ minimum as a function of $\ep$.
Writing it as $\ch^2_\ep$, the fit quality is measured as $\ch^2_\ep/$dof, where $\text{dof}=1$ in our analysis.

The extension to the case of more than three isotope pairs or more than two transitions is straightforward.

\end{document}